\documentclass[aps,preprint]{revtex4}

\usepackage{color}
\usepackage{epsfig}
\usepackage{amsmath}
\usepackage{epstopdf} 
\usepackage{mathrsfs}
\usepackage{graphicx}
\usepackage{ulem}

\begin{document}

\title{Fluorescence decay enhancement and FRET inhibition in self-assembled hybrid gold CdSe/CdS/CdZnS colloidal nanocrystals supraparticles}

\author{V. Blondot,$^{1}$ C. Arnold,$^{1}$ A. Delteil,$^{1}$  D. G\'{e}rard,$^{1}$, A. Bogicevic,$^{2}$ T. Pons,$^{2}$ N. Lequeux,$^{2}$ J.-P. Hugonin,$^{3}$  J.-J. Greffet,$^{3}$ S. Buil, $^{1}$ and J-P. Hermier $^{1,\ast}$}

\author{$^{1}$ Universit\'e Paris-Saclay, UVSQ, CNRS, GEMaC, 78000, Versailles, France.\\
$^{2}$ Laboratoire de Physique et d'\'Etude des Mat\'eriaux, ESPCI-Paris, PSL Research University, CNRS UMR 8213, Sorbonne Universit\'e, 10 rue Vauquelin, 75005 Paris, France\\
$^{3}$ Universit\'e Paris-Saclay, Institut d’Optique Graduate School, CNRS, Laboratoire Charles Fabry, F-91127
Palaiseau, France}

\author{$^{\ast}$ jean-pierre.hermier@uvsq.fr}

\begin{abstract}
We report on the synthesis of hybrid light emitting particles with a diameter ranging between 100 and 500 nm, consisting in a compact semiconductor CdSe/CdS/CdZnS nanocrystal aggregate encapsulated by a controlled nanometric size silica and gold layers. We first characterize the Purcell decay rate enhancement corresponding to the addition of the gold nanoshell as a function of the particle size and find a good agreement with the predictions of numerical simulations. Then, we show that the contribution corresponding to F\"orster resonance energy transfer is inhibited.
\end{abstract}

\maketitle

Plasmonic structures enable to modify the photoluminescence properties of individual fluorophores for a wide range of applications such as quantum optics \cite{Tame13,Zhou19}, optoelectronics \cite{Liang14} or the detection of molecules of chemical or biological interest \cite{Brolo12}. When colloidal semiconductor nanocrystals (NCs) with a CdSe core are considered, many configurations involving plasmonic surfaces or nanoparticles were reported \cite{Gomez10,Hoang15,Yao17,Coste20}. They allow to enhance the emission intensity, to control the emission pattern or to suppress the NC residual blinking. 

From a general point of view, as an alternative to photonic structures, confinement of multiple emitters in a small volume offers the possibility to benefit from their interaction in order to enhance the emission through coherent dipole-dipole interactions leading to superradiance or superfluorescence \cite{Dicke54}. Plasmonic devices and cooperative effects can be combined and hybrid metal/emitters structures were designed to develop SPASER for example \cite{Azzam20}.

Dipole-dipole interactions can also generate a non-reversible and incoherent energy transfer which corresponds to FRET (F\"orster resonance energy transfer), especially when condensed matter nanoemitters showing a large inhomogeneity of their spectral properties such as NCs are considered \cite{Anni04,Jolene14}. Widely used for the investigation of biological or chemical systems \cite{Medintz03,Pons06}, FRET and its control through optical cavities is also the subject of numerous works in the field of optoelectronics for the development of devices such as LEDs or photovoltaic devices \cite{Zhang20,Li15,Hsu17}. On a fundamental level, C. L. Cortes and Z. Jacob have shown that the fluorescence enhancement by the Purcell effect could be tuned independently of the increase or suppression of FRET \cite{Cortes18}. 

In a recent paper, we have demonstrated the synthesis of NC aggregates surrounded by a silica shell, called supraparticles (SPs) in the following, which exhibit a perfectly stable emission intensity at room temperature \cite{Blondot20}. This study also highlighted that the decay rate corresponding to FRET is of the order of the radiative decay rate. Here, we characterize the emission properties of similar structures when coupled to a nanoresonator consisting of a thin gold shell encapsulating the SPs. We first describe the synthesis of these nanoobjects and their elementary structural properties. Next, using several spectroscopic and/or complementary time resolved setups, we show the reduction of the FRET efficiency induced by the gold nanoshell. 

\section{Chemical synthesis and structural properties of GSPs}
A multistep procedure was performed to prepare a hybrid structure containing few hundred NCs inside a gold nanoshell.
\subsection{Superparticles synthesis}
First, CdSe NCs were synthesized by a procedure slightly modified from Cao et al.  \cite{Cao} and shells of CdS/CdZnS were synthesized with continuous injection of shell precursors at high temperatures \cite{Boldt2013}. Typically, these reactions generate CdSe/CdS/CdZnS NCs of about 7.7 ± 1 nm in size with an emission centered at 645 nm (FWHM = 30 nm). These NCs are then assembled into spherical aggregates using a micro-emulsion/evaporation method  \cite{Bai2007}. Typically, 6 nmol of NCs were re-dispersed in 1 mL of chloroform and added to 1 mL of dodecyltrimethylammonium bromide solution (20 mg/mL in water). The solution was vortexed for 30 seconds to obtain a stable oil in water emulsion. The chloroform was evaporated by heating the solution at 70°C in an oil bath for 10 minutes. The obtained NC assemblies stabilized by DTAB are hereafter referred to as NC supraparticles, SPs. They were purified by centrifugation at 6 000 g for 5 minutes and re-dispersed in 4 mL of ethanol. NC SPs were coated with a silica shell (20 nm in thickness to prevent thereafter the quenching by the gold nanoshell of the emission of the NCs located close to the surface of the SP) by hydrolysis and condensation of tetraorthoethylsilicate (TEOS) precursors.  For a typical reaction, 1 mL of NC SP solution was first sonicated in 9 mL of ethanol and 1.5 mL of deionized water for 10 minutes. Then 0.5 mL of ammonia solution (NH4OH, 28 wt\% in water) and 10 µL TEOS were added. The mixture was sonicated for 20 minutes to prevent aggregation \cite{Ammar2007}. The silica-coated NC SPs were purified by centrifugation and redispersed 1 mL of ethanol. Silica shells were functionalized by reaction with aminosilane precursors \cite{Villa2016}. Typically, 500 µL of NC SPs coated with silica were added to a mixture of 500 µL of ethanol, 1 mL of water and 50 µL of aminopropyltriethoxysilane and reacted at 60°C for 2 hours. The solution was washed 3 times with ethanol and twice with water to remove excess silane and then re-dispersed in 1 mL of water. The growth of gold nanoshells was performed using a protocol derived from Halas and coworkers \cite{Oldenburg1998}. Gold seeds were prepared by the Duff method \cite{Duff1993}. 100µL from the previous 1 mL functionalized NC SPs were mixed with 1 mL of gold seeds solution for 30 minutes. The NC SPs were purified by centrifugation in water and re-dispersed in 300 µL of water. 100 µL from this solution were then dispersed in 5 mL of gold growth solution \cite{Oldenburg1998} followed by the addition of 25 µL of formaldehyde and stirred for 2 h. The final gold-coated NC SPs (golden supraparticles, GSPs) were centrifuged in water and re-dispersed in 500 µL of water. The full description and characterization of the NC GSP synthesis at the different steps of the preparation are described in a separate report \cite{Bogicevic2022}.

\begin{figure}[htp]
\centering
\includegraphics [width=14cm]{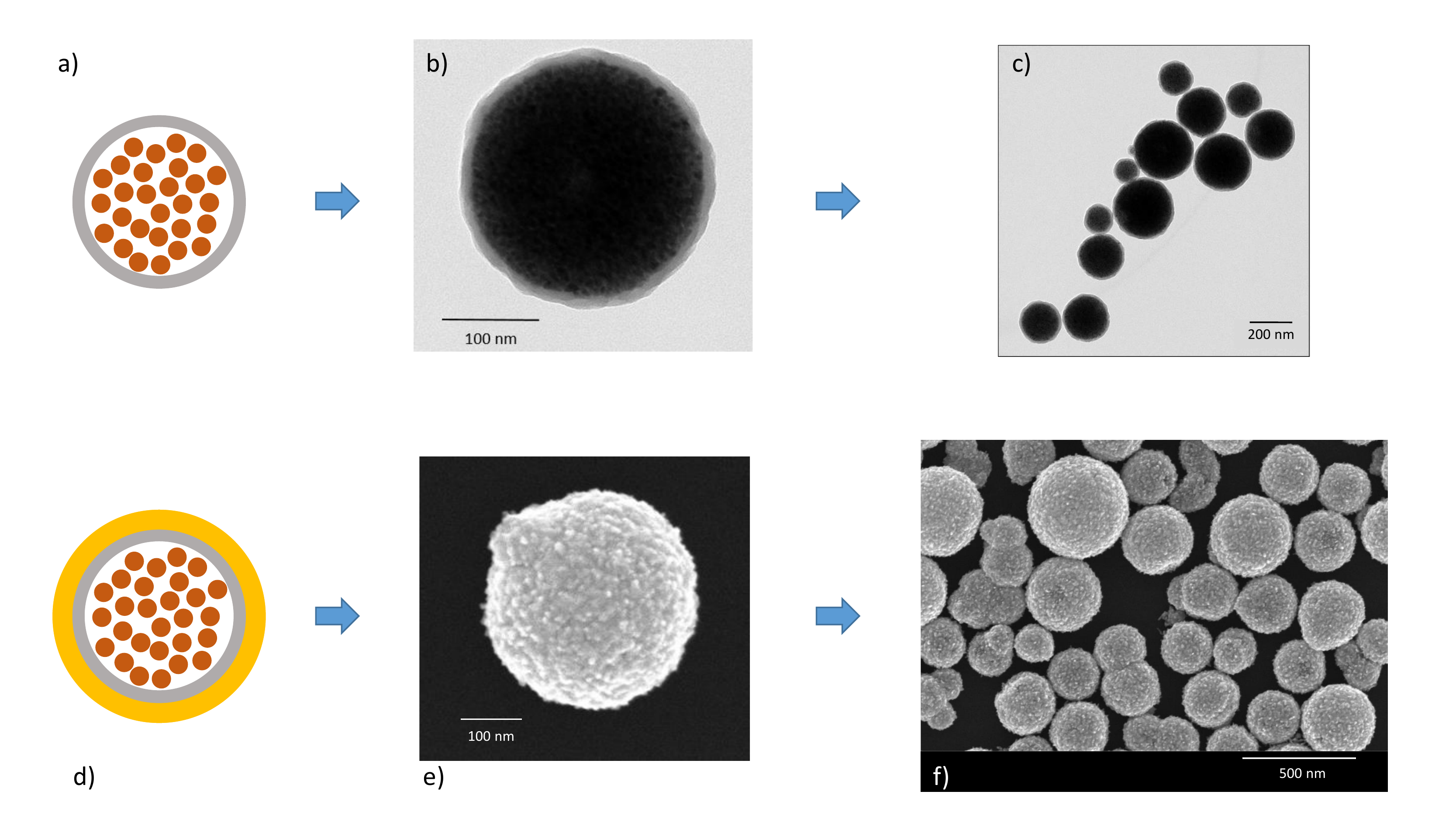}
\caption{Schematic representations of a single supraparticles of NCs after coating with silica (a) and gold (d). Transmission electron microscopy image of a single SP (b) and Scanning electron microscopy image of a GSP (e). Large field of view of images for SPs (c) and GSP (f).}\label{figure_1}
\end{figure}

\begin{figure}[htp]
\centering
\includegraphics [width=14cm]{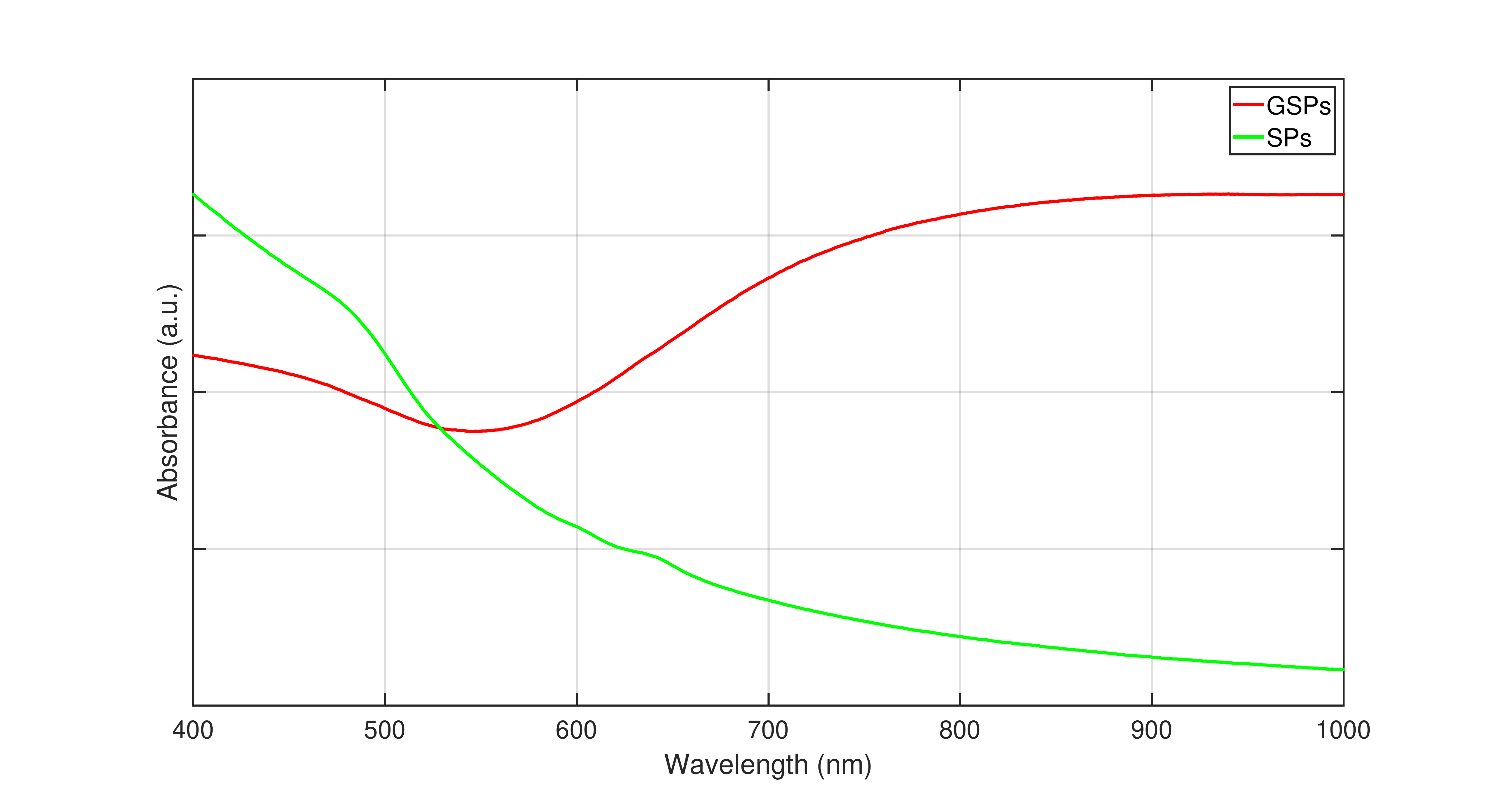}
\caption{Absorbance of SP and GSP solutions.}\label{figure_2}
\end{figure}

\subsection{Gold nanoshell deposition}
We follow the common synthesis route developed by Halas and coworkers \cite{Oldenburg1998} to obtain gold nanoshells with a thickness of 30 ± 4 nm. First, silica layer must be functionalized by organosilane molecules (3-aminopropyltriethoxysilane, APTES) to generate a new surface termination and allow the gold seeds deposition.  A solution containing 2-3 nm gold seeds prepared according to Duff {\it et al.} \cite{Duff1993} is added to APTES functionalized SPs and the solution is gently stirred for an hour. Several washing steps with water are necessary to remove excess of gold seeds in solution. Gold seeds immobilized onto the silica layer act as nucleation sites for electroless Au plating using an aqueous HAuCl$_4$ plating solution in the presence of formaldehyde \cite{Brinson2008, Wang2006}. This last stage consists of the growth and coalescence of gold seeds to form a continuous gold nanoshell onto SPs (see Fig. \ref{figure_1}.d to \ref{figure_1}.f), named hereafter golden supraparticles (GSPs). The addition of the gold nanoshell results in a strong absorption above 600 nm corresponding to a large plasmon band (see Fig. \ref{figure_2}) \cite{Brinson2008,Zhou94}

\section{Emission properties}
\subsection{Experimental setup}
To characterize the emission properties of SPs at the single object level, we spin coat the solution of the particles on a glass coverslip. In the case of the GSPs, the concentration is so low that we only deposit 40 $\mu$l of the solution and let the solvent evaporate. A copper electron microscopy grid of 32 numbered cells of 10 $\mu$m $\times$ 10 $\mu$m is then glued to the coverslip (Gilder Finder Grids G400F1, SigmaAldrich). Using an atomic force microscope (AFM), we locate the individual particles with respect to the grid and measure their size so that the confocal experiments will be performed on supraparticles whose size are known.
\begin{figure}[htp]
\centering
\includegraphics [width=14cm]{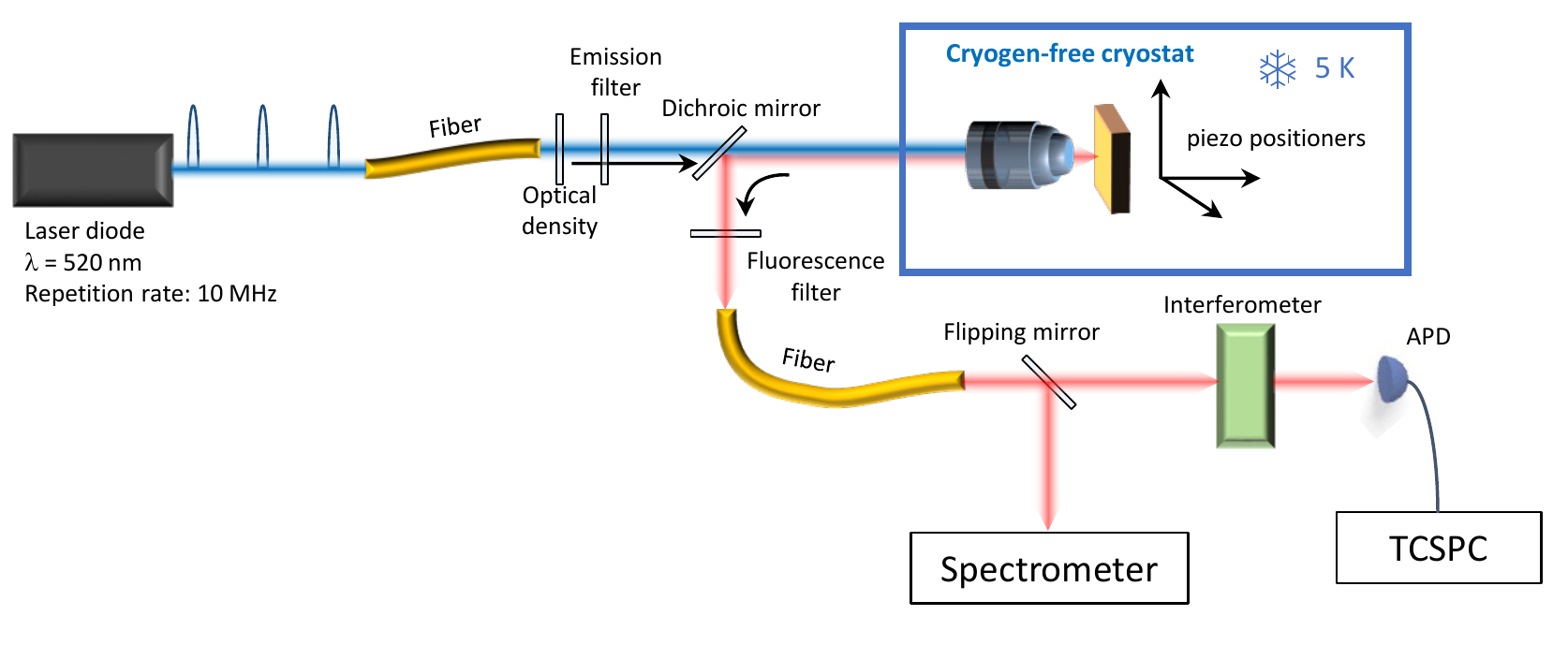}
\caption{Set-up for time and spectrally resolved photoluminescence experiments.}\label{figure_3}
\end{figure}

In order to carry out the optical measurements (see Fig. \ref{figure_3}), a confocal microscope (Attocube Attodry 1100, numerical aperture of the objective = 0.82) operating at 4 K (Cryomech cryocooler) focuses a pulsed laser diode beam (Picoquant, LDH-D-C-520, FWHM $\sim$ 160 ps) of low fixed power (the probability to excite one NC of the aggregate is of the order of 10 \%) at the diffraction limit on particles that were previously identified with the AFM. Photons are detected with single photon avalanche photodiodes (SPAD) (MPD, time resolution 50 ps) connected to a Time Correlated Single Photon Counting (TCSPC) device (Picoquant, PicoHarp 300) that enables to measure the photoluminescence decay. An interferometer (GEMINI, Nireos model) can be placed before the SPAD, which provides after a Fourier transform the fluorescence as a function of detection wavelength and the photon detection time (maximal spectral resolution $\sim$ 0.7 nm). Finally, a spectrometer and a cooled CCD camera allow to reduce the spectral accuracy down to 0.05 nm, at the cost of a time resolution of several seconds.

\subsection{Emission of supraparticules and Purcell effect}
\begin{figure}[htp]
\centering
\includegraphics [width=14cm]{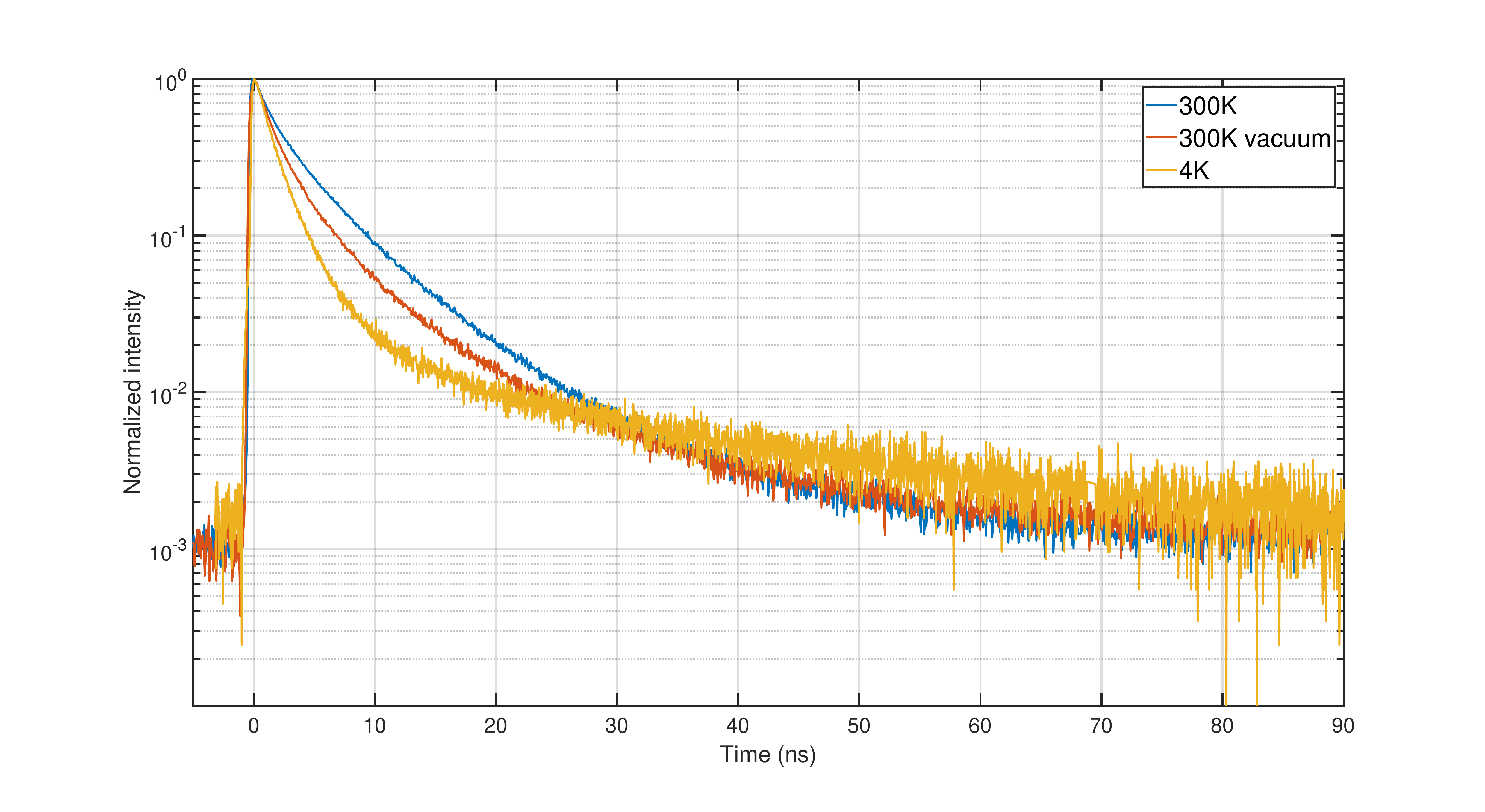}
\caption{Photoluminescence of a single GSP in air and at 300 K (red), in vacuum and at 300 K (blue) and at 4 K (yellow).}\label{figure_4}
\end{figure}

Fig. \ref{figure_4} shows the PL decay of a typical GSP under low excitation power and under various conditions: at 300 K in air or vacuum (low helium pressure $\sim 10^{-3}$ mbar), or 4 K (same low helium pressure). In vacuum, thick shell CdSe/CdS NCs are well known to become ionized since no water or oxygen molecules can neutralize the emitter once it has been photoionized \cite{Javaux13}. The photoluminescence (PL) lifetime then decreases since the trion recombines twice as fast as the exciton \cite{Galland12}. For the single CdSe/CdS/CdZnS NCs used to synthesize the SPs and GSPs, we measured a PL lifetime about 5.6 ns in vacuum. As a result, the increase of  the PL decay rate of GSPs at 300 K when placed under vacuum compared to in air (Fig. \ref{figure_4}) shows that NCs inside a GSP become ionized. Indeed, the porosity of the chemically synthesized silica and gold shells does not prevent the escape of the water or oxygen molecules from the SPs or GSPs. In the following, the experiments are performed at 4 K and we will consider that the NCs forming the aggregates are always ionized. 

If we consider a single NC inside an aggregate, the close proximity of the surrounding NCs composed of materials with refractive indices around 2.5 is likely to modify its electromagnetic environment compared to single NCs in vacuum. Moreover, the sub-wavelength dimension of the SPs results in the confinement of the electromagnetic field, leading to a Purcell-enhanced emission. In order to investigate the influence of the surrounding environment for NCs inside the SPs, we use a multilayer spherical model based on Mie theory. We model the aggregate part of our structures as a homogeneous material with a permittivity being the average of the permittivities of the different materials composing the NCs weighted by their relative fraction in the total volume. We assume a packing density of NCs of $66\%$ in the aggregates \cite{Zaccone22}. We point out that this model can give information on Purcell effect but does not account for dipole-dipole interactions between NCs. Therefore, FRET is put aside in this first analysis.

To model gold-coated aggregates, the roughness of the metallic layer is described by a homogeneous layer following the method of ref. \cite{botao15}. The gold refractive index is defined thanks to a Drude model adjusted from ref. \cite{Johnson72}. To account for the specific optical properties of gold obtained by chemical process, we followed ref. \cite{botao15} where a fivefold increase of the optical losses and a slight adjustment of the plasma frequency is done on the Drude model. These high Joule losses result in plasmonic modes which show no resonances. 

	Using Mie theory, we expand the field emitted by a single dipole on a basis of spherical harmonics. We can then compute the power emiited by the dipole inside the supraparticle \cite{Chew87}. Transition rates of atoms near spherical surfaces] and average over all wavelengths and over the different positions of the dipole inside the aggregate. From the emitted power, we can derive the decay rate.  In principle, we should also take into account the dielectric interface corresponding to the glass coverslip that modifies a dipole emission located at a distance lower than typically 0.1 $\lambda$ = 60 nm \cite{Chance1974,Lukosz77, Egel2016}. However, calculations show that the interface only slightly modifies the radiated power, which is consistent with the size of the SPs (greater than 200 nm) and the shell thickness (20 nm for the SPs and 50 nm for the GSPs). Finally, to calculate the Purcell factor, we use a single dipole above a glass substrate as reference (the latest is obtained following \cite{Chance1974}).

\begin{figure}[htp]
\centering
\includegraphics [width=14cm]{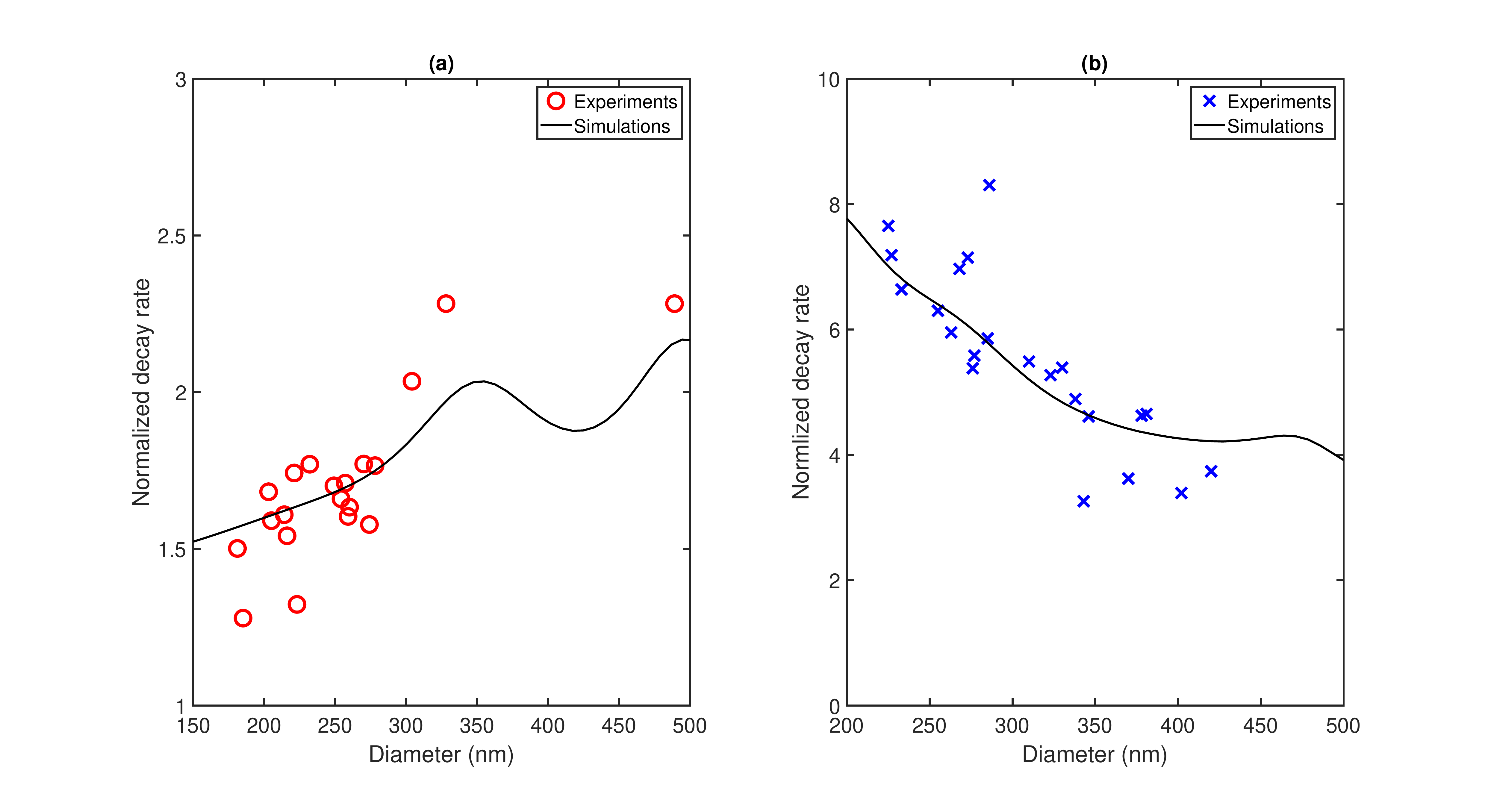}
\caption{Decay rates of SPs (a) and GSPs (b) normalized by the decay rate of individual emitters. Black curves correspond to the decay rates obtained by the numerical simulations. Crosses and circles are the decay rates of the log-normal distribution used to fit measurements.}\label{figure_5}
\end{figure}

Experimentally, we first fitted the PL decay rates of single NCs at 4K under vacuum was first fitted by a bi-exponential function, the first component corresponding to the decay of the trion \cite{Javaux13}. The luminescence decays of individual aggregates are fitted with a log-normal distribution with a central decay rate $\Gamma_{LN}$ \cite{Blondot20} for the NC emission plus a mono-exponential function to take into account a slow emission from traps \cite{Sher08}. Fig. \ref{figure_5}.a shows the decay rates of non-gold coated SPs as a function of their diameters, normalized by the mean decay rate of single NCs deposited on a glass substrate. A good agreement is obtained between the measured decay rates $\Gamma_{LN}$ and the decay rates predicted by the model. Emission enhancement is measured between 1.3  and 2.3  in comparison with single NCs. These accelerations come from the high refractive index of the aggregate as well as  Mie resonances. 

We now turn to decay rates of gold-coated GSPs (Fig. \ref{figure_5}.b). As expected, the emission enhancement is even more pronounced than for non-metallized aggregates, particularly for the smaller GSPs where accelerations of 8 compared to individual NCs have been obtained. 

It should be noted that the good agreement between model predictions and measurements for GSPs and SPs was not obvious, given the FRET is not taken into account in our calculations. However, if the quantum efficiency and the radiative decay rate ''blue'' NCs (corresponding to the NCs with the shortest wavelength emission) are the same as the "red" ones, the FRET does not modify the overall luminescence decay time. Indeed, let us consider the basic situation of excited donor (D) and acceptor (A) emitters characterized by single fluorescence ($k$) and  FRET ($k_{FRET}$) decay rates. This simplifed system can be depicted by the following rate equations:
 
\begin{eqnarray*}
\frac{d N_D}{dt} = - k_{FRET} N_D - k N_D \\
\frac{d N_A}{dt} = +k_{FRET} N_D - k N_A 
\end{eqnarray*}

By summing the two equations, one finds that the total number of excited emitters $N=N_D+N_A$ follows:

\begin{equation*}
\frac{d N}{dt} = - k N
\end{equation*}

that means that the overall decay rate is left unchanged. The generalization to multiple population classes is straightforward and leads to the same conclusion. The FRET only modifies the spectrum that is redshifted. This first analysis focused on Purcell effect, to compare FRET between NCs in the SPs and GSPs. In the next section, we turn into the measurements allowing spectrally-resolved measurements of the decay rates. 

\subsection{FRET}
\begin{figure}[htp]
\centering
\includegraphics [width=14cm]{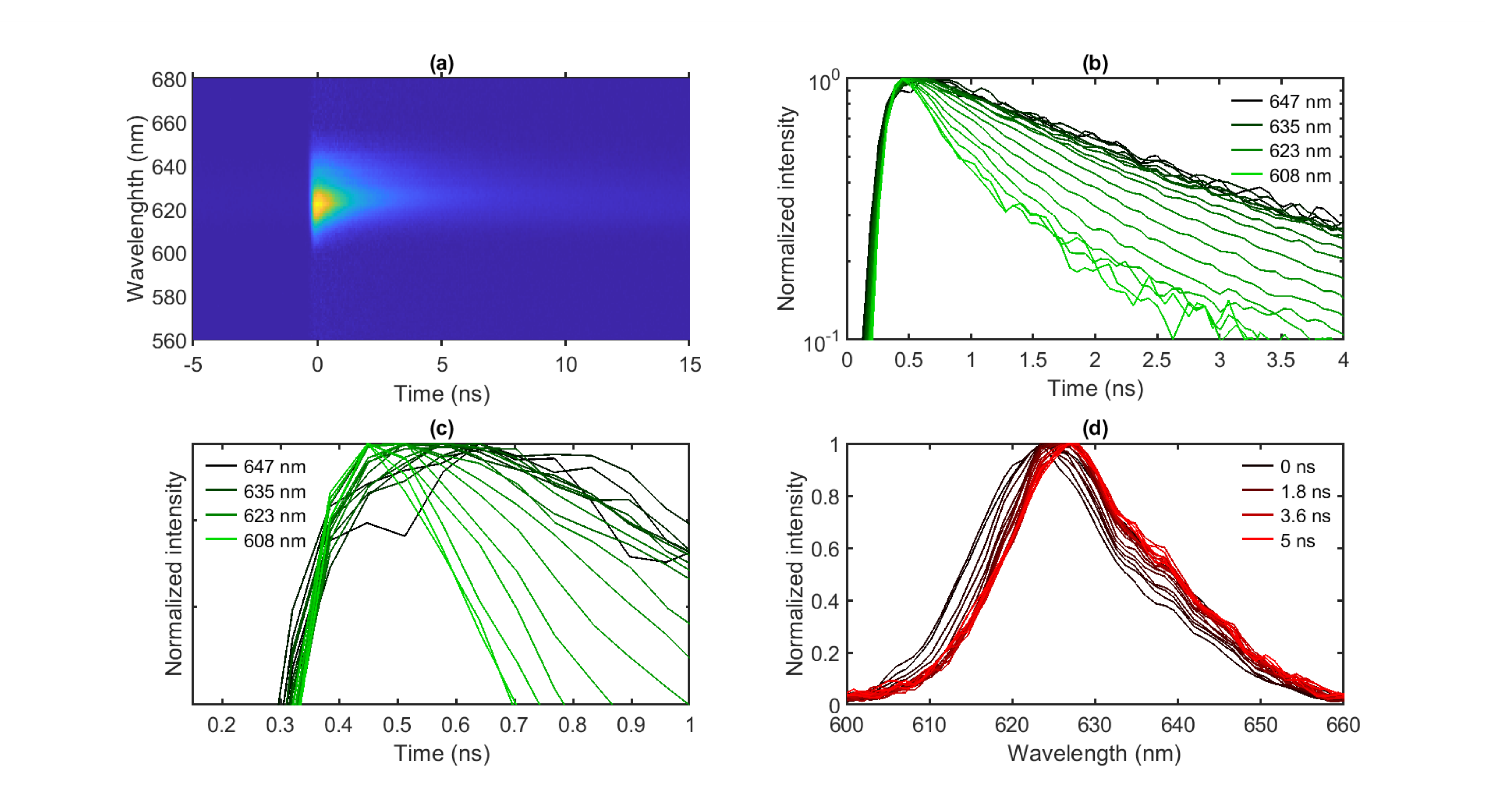}
\caption{(a): example of a fluorescence map of a single SP. (b): PL decay for a SP as a function of the wavelength from 647 nm (black curve) to 608 nm (lightest green). (c): zoom of the same PL decay. (d): Emission spectrum of the same SP as a function of time from 0 ps (black curve) to 5 ns (lightest red).}\label{figure_6}
\end{figure}

Aggregates of NCs are known to exhibit FRET \cite{Jolene14,Medintz03,Pons06,Rafipoor18}. In a SP, the energy transferred from a NC is all the greater as its wavelength emission is blueshifted, as a result of an increased overlap between its emission spectrum and the average absorption spectrum of its neighbours \cite{Blondot20}. We now show that the gold nanocavity modifies the impact of FRET on the PL emission properties of the GSPs with respect to the SP ones. As already mentionned, the Fourier transform of the signal recorded by the SPAD at the ouput of the interferometer enables to plot a fluorescence map as a function of the detection wavelength and the time of photon detection $\tau$ (Fig. \ref{figure_6}.a). From this data, we can plot the PL decay for a given wavelength (Fig. \ref{figure_6}.b) or, alternatively, the evolution of the emission spectrum at a given time after the pulse excitation (Fig. \ref{figure_6}.d). Both data sets demonstrate the strong effect of FRET on non-metallized SPs. First, PL decays for shorter wavelengths are faster than for longer ones.  FRET is also evidenced by the wavelength dependence of the delay between the pulse excitation and the PL signal peak. Due to energy transfer from the "blue" NCs to the "red" ones, the maximum emission of the "red" is delayed with respect to the "blue" one. Lastly, the normalized spectrum is redshifted over time (Fig. \ref{figure_6}.d). 

\begin{figure}[htp]
\centering
\includegraphics [width=14cm]{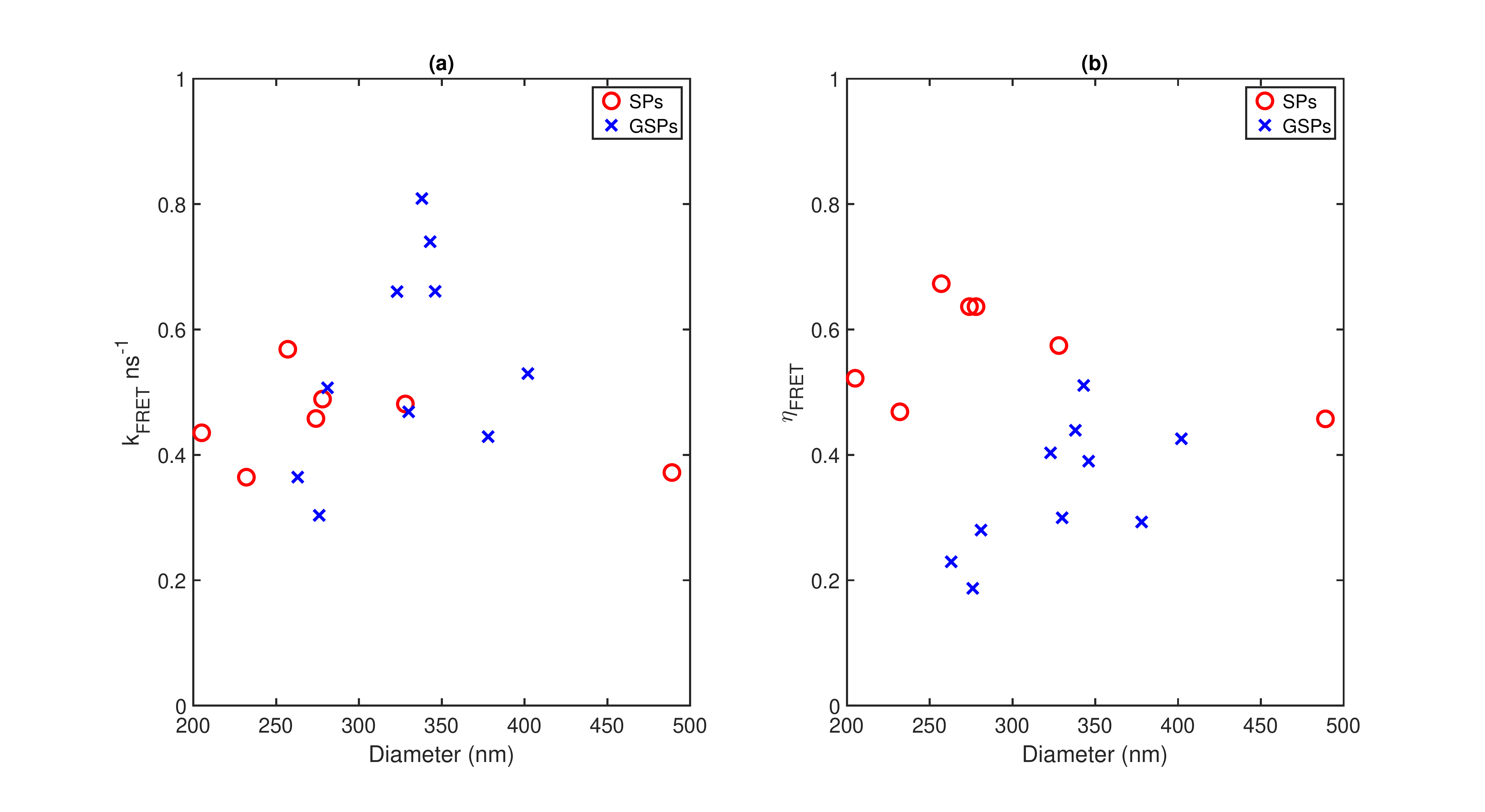}
\caption{FRET analysis on SPs (red circles) and GSPs (blue crosses). (a): FRET decay rates obtained by doing the differences of the decay rates at 615 and 640 nm as a function of the diameters. (b): FRET efficiencies, defined as the ratio between $k_{FRET}$ and the decay rates measured at short wavelengths.}\label{figure_7}
\end{figure}

Simulations done on SPs show negligible variation of the Purcell-enhanced decay rates when the wavelength varies inside the  NCs emission spectrum (not shown). Since the decay rate of the trion $k$ does not depend on the NC emission wavelength \cite{Blondot20}, the variations of the decay rate with the wavelength can be entirely attributed to FRET. As a result, the decay rate $k(\lambda)$ for NCs inside a GSP emitting at a wavelength $\lambda$ writes:

\begin{equation*}
k(\lambda)=k+k_{FRET}(\lambda)
\end{equation*}

where  $k_{FRET}(\lambda)$ is the decay rate due to FRET. The variations that we measure between the fast decay rates of the "blue" NCs and the slower "red" NCs comes from the FRET decay rate $k_{FRET}$. In \cite{Blondot20}, we have also shown that the energy transferred by FRET from a given NC decreases with its wavelength emission and vanishes for the highest wavelength, that is $\lambda =647$ nm. Fig. \ref{figure_7}.a shows the value of $k_{FRET}$ deduced by the difference between the decay rates measured at 640 nm and at 615 nm.  An average value of $k_{FRET}(\lambda = 615 nm)= 0.45$ ns$^{-1}$ is obtained (spectral filtering reducing the number of counts used to plot the PL decay, we used a basic monoexponential fit). The variations between SPs can be due to the difficulty to adjust the weak luminescence decays at the edges of the emission spectrum.

\begin{figure}[htp]
\centering
\includegraphics [width=14cm]{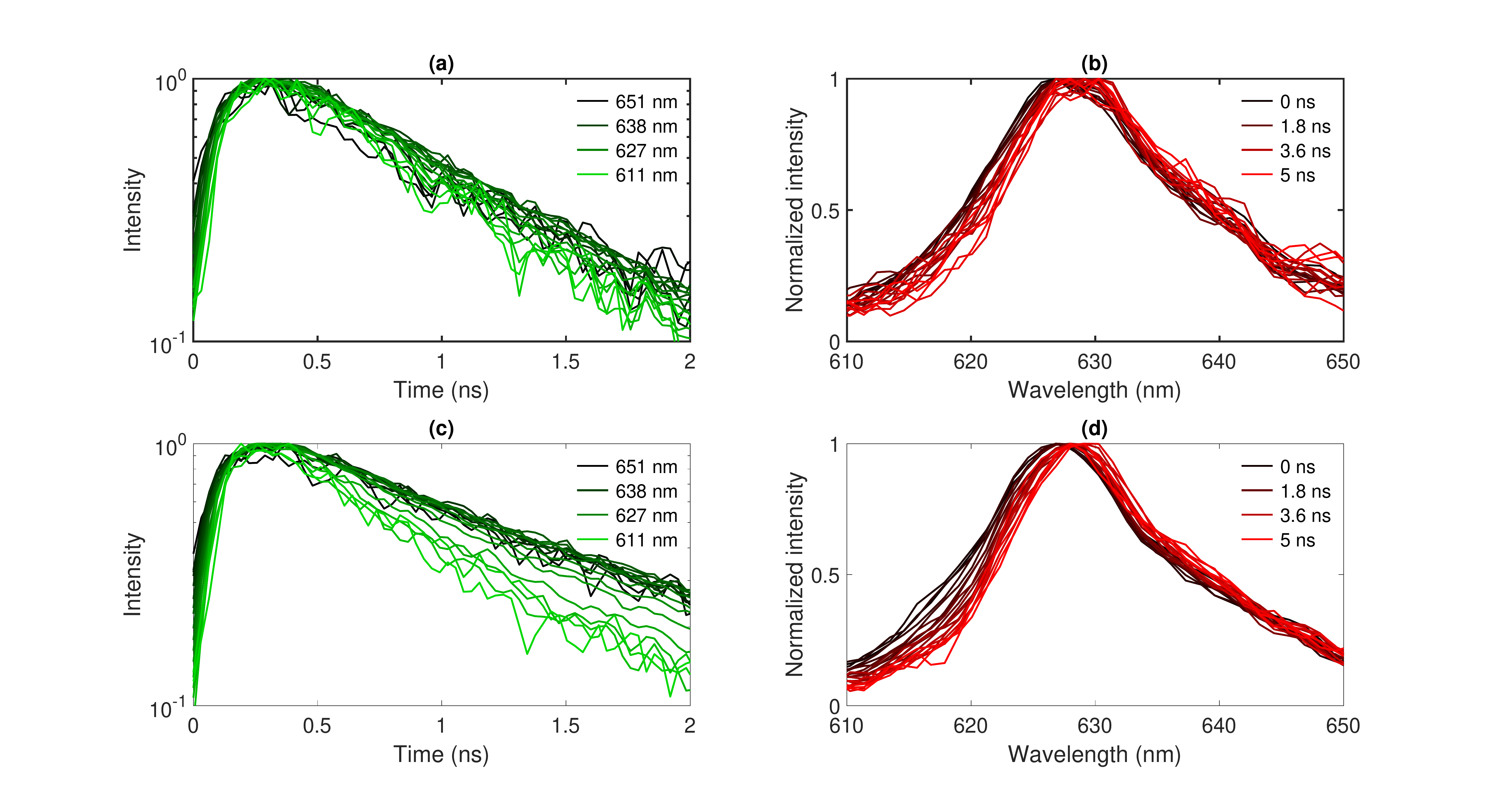}
\caption{(a): PL decay for GSP$_1$  as a function of the wavelength from 651 nm (black curve) to 611 nm (lightest green). (b): Emission spectrum of  GSP$_1$  as a function of time from 0 ps (black curve) to 5 ns (lightest red). (c): PL decay for GSP$_2$  as a function of the wavelength from 651 nm (black curve) to 611 nm (lightest green). (d): Emission spectrum of  GSP$_2$  as a function of time from 0 (black curve) to 5 ns (lightest red).}\label{figure_8}
\end{figure}

In contrast with SPs, the effect of FRET is far less visible for the GSPs. The differences between the decay rates of ''blue'' NCs and ''red'' NCs are less marked (Fig. \ref{figure_8}.c) than on SPs and even hardly visible on some GSPs (Fig. \ref{figure_8}.a). This is explained by the enhancement of the emission by Purcell effect which dominates over the FRET rates. We also point out that the shift of the PL decay maximum (Fig. \ref{figure_6}.c) with the wavelength is no more visible.  Simultaneously, the modification of the normalized spectrum over time is small (Fig. \ref{figure_8}.b and Fig. \ref{figure_8}.d, to be compared with Fig. \ref{figure_6}.b and Fig. \ref{figure_6}.d), and all the more so as the contribution of FRET is low. Fig. \ref{figure_7}.a shows that we obtain comparable $k_{FRET}$ on GSPs than on SPs. As shown by C. L. Cortes and Z. Jacob in their detailed modeling \cite{Cortes18}, the effect of the gold plasmonic resonator on the FRET rate is not straightforward: it can either have no effect, increase or decrease the ratio between the total decay rate and the FRET rate in a way that depends strongly on the position of the donor-acceptor pair with respect to the photonic mode, independently of the evolution of the radiative rate. 

In order to determine the probability that a NC decays by FRET rather than other recombination channels, we deduced from measurements the FRET efficiency which is defined as the ratio between $k_{FRET}$ and the decay rates measured at short wavelengths (Fig. \ref{figure_7}.b). For the smallest structures (below 300 nm), the FRET efficiency of GSPs is reduced by a factor of 3 compared to non-metallized SPs, showing the strong impact of the metallic layer to reduce FRET of NCs inside SPs. We can also notice that FRET efficiencies become comparable for larger diameters, which was expected since the Purcell effect becomes similar for SPs and GSPs when the size increases. 

\begin{figure}[htp]
\centering
\includegraphics [width=14cm]{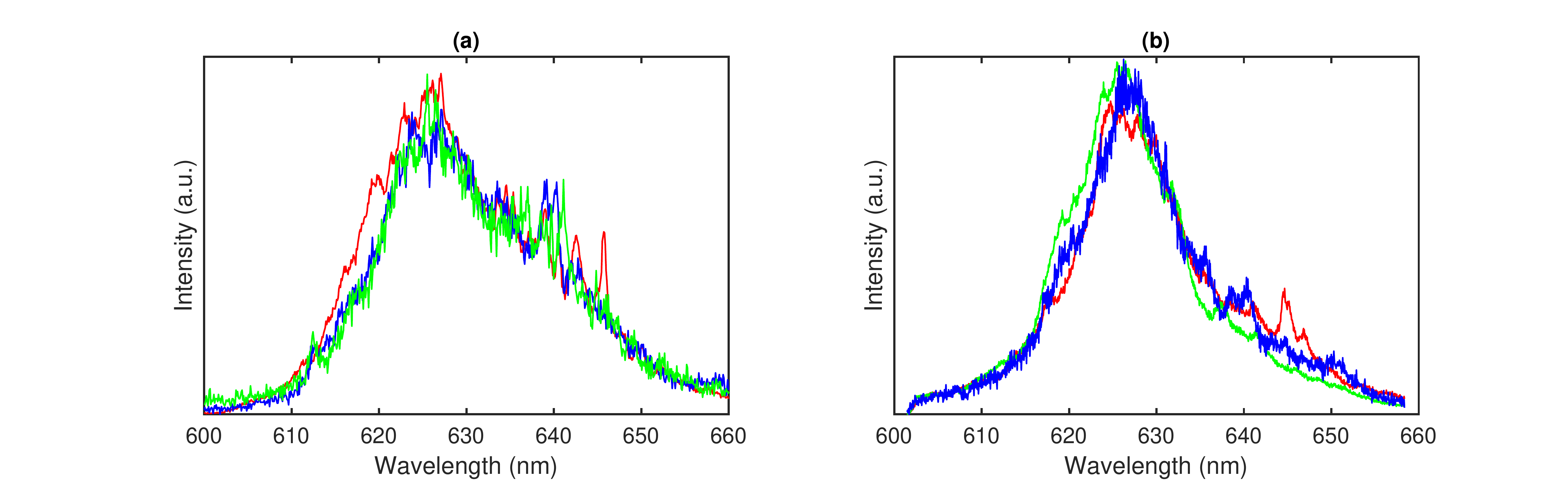}
\caption{Emission spectrum of three single SPs (a) and three single GSP (b).}\label{figure_9}
\end{figure}

The modification of the FRET efficiency is also confirmed by the shape of the emission spectrum integrated over 10 s. The FRET broadens the overall spectrum for the SP with respect to the GSP one (Fig. \ref{figure_9}). In the case of SP, FRET favors the emission of the "red" NCs through their excitation by the "blue" ones. The result is a relative increase of the "red" part of the emission spectrum and a relative decrease of the central peak height (the systematic asymmetry of the emission spectrum towards the red indicates that a small fraction of the NCs exhibit a slightly larger size). 

\section{Conclusion}
In conclusion, we first have presented the synthesis of hybrid structures consisting in NCs aggregates surrounded by a gold nanoresonator. As expected, the PL decay rate is increased by the metallic structure. We have shown that the mean Purcell factor is well predicted by a model and depends on the particle size. In addition, through the study of the SPs with or without a gold shell, we have also analyzed in detail the influence of the metallic resonator on FRET. Especially, interferometric measurements providing time resolved fluorescence spectra show that the relative contribution of FRET decreases by a factor of 3 for the smallest GSPs. Beyond the fine structure of the PL temporal dynamics, we have also found that the overall spectrum of the GSPs is narrowed with respect to the SPs ones. In terms of applications, the GSPs are submicronic bright and photostable sources that could be used for example for lighting. More fundamentally, the inhibition of FRET, which is an incoherent energy transfer process, opens the possibility to achieve coherent interactions between NCs that could enable to achieve quantum collective emission regime such as superradiance.

\section{Acknowledgments}
This work is supported by the Agence in the framework of the GYN project (Grant No. ANR-17-CE24-0046).

\newcommand{\enquote}[1]{``#1''}

\end{document}